\documentclass[a4paper,11pt]{article}
\pdfoutput=1 
\usepackage{jcappub} 
\usepackage{graphicx}
\usepackage{natbib}
\begin{document}

\title{Efficiently Extracting Energy  from Cosmological Neutrinos}
\author{M.M. Hedman}
\affiliation{Center for Radiophysics and Space Research, Cornell University, Ithaca NY 14853}
\affiliation{Department of Physics, University of Idaho, Moscow ID 83844-0903}
\emailAdd{mhedman@uidaho.edu}
\date{\today}

\abstract{
Detecting the extremely low-energy neutrinos that form the Cosmic Neutrino Background  (CNB) presents many experimental challenges, but pursuing this elusive goal is still worthwhile because these weakly-interacting particles could provide a new window into the structure and composition of the early universe. This report examines whether cosmological neutrinos can deposit sufficient energy into a target system to be detectable with plausible extensions of current bolometric technologies. While the macroscopic wavelengths of cosmological neutrinos can greatly enhance their cross sections with dense targets, such interactions can only be detectable if they transfer a significant fraction of each neutrino's kinetic energy into the detector system. We find that a large array of dense target masses coupled to suitable motion-sensitive circuits could potentially satisfy both of these conditions and thus might be able to serve  as the basis for a more practical cosmological neutrino detector.}



\maketitle

\section{Introduction}

As the universe cooled down from its hot and dense  initial state, it underwent a series of phase transitions where particle species that were initially in thermal equilibrium decoupled from each other. The most relativistic of these particles   have been free-streaming through the universe until the present day, and therefore can carry information about the state of the early universe to detectors currently operating on Earth. The most famous of these relic particle populations is the Cosmic Microwave Background, or CMB, which is one of today's most powerful cosmological probes (see e.g \cite{Komatsu11, Keisler11, Hinshaw12, Planck13a}) However, the CMB is not the only cosmological particle background in the universe. In particular, there should also be a Cosmic Neutrino Background, or CNB.

Standard cosmological models predict the present universe is filled with a  nearly homogeneous population of each of the three neutrino species and three antineutrino species. These neutrino populations should follow a  thermal Fermi-Dirac distribution with an effective temperature of 1.95 K,  such that the mean number  density of each species of neutrino or anti-neutrino is  $\sim 56/$cm$^3$ \citep{Peebles93}. Recent cosmological observations, including measurements of the CMB power spectrum, are roughly consistent with this basic model \citep{Keisler11, Hinshaw12, Planck13b}, and thus give us confidence that such a background does indeed exist. If these cosmological neutrinos could be observed, they could provide invaluable information about the structure, composition and early history of the universe \citep{MC07, DV09}. 

Directly detecting these extremely low-energy and  weakly-interacting particles poses major experimental challenges, and while a variety of detection techniques have been considered over the last four decades, none has proven to be practical with reasonable extensions of current technologies. For example, multiple authors have explored the possibility of detecting the mechanical forces cosmological neutrinos can apply to various laboratory systems \citep{Stodolosky75, CM82, Shv82, SL83, Langacker83,  Stodolsky89, FW95, Duda01, Gelmini05, Ringwald09}. These scenarios often rely heavily on enhancement factors in the interaction cross sections that arise when the low-energy neutrinos coherently scatter from all the particles within their de Broglie wavelength. However, even with the resulting enhancements in the interaction rates, the resulting accelerations are still orders of magnitude below the detection threshold of any laboratory system conceived thus far \citep{CM82, Shv82, Stodolsky89, FW95, Duda01, Ringwald09}. 

More recently, there has been interest in the possibility of detecting cosmological neutrinos in beta-decay experiments such as KATRIN and MARE \citep{Kaboth10, Betts13}.  In these cases the observational signature is a distortion in the energy spectra of the electrons emitted by  unstable nuclei that arises because some of these nuclei capture a cosmological neutrino instead of undergoing normal beta decay. Compared with detecting mechanical forces, these techniques have the advantage that each neutrino interaction produces a distinct change in the target system. Unfortunately, the neutrino cannot be coherently captured by many nuclei, so the relevant reaction cross sections are not enhanced by the neutrinos'  macroscopic wavelengths.  Indeed, the rates of such relic neutrino captures are so low that planned experiments like KATRIN could only detect a signal from the CNB if it was $\sim10^9$ times  larger than that expected in standard cosmological models \cite{Kaboth10}. 
Thus, while these machines could detect non-standard neutrino backgrounds, they may not yet provide a clear path towards a practical method of detecting the CNB expected from standard cosmological models. 

Despite these discouraging findings, one must remember that mechanical forces and nuclear transitions are only two ends on the spectrum of detectable signals that could be produced by cosmological neutrinos. In between these limiting cases are a wide range of  detection schemes where the neutrinos excite transitions between internal states of the target system. These sorts of detection scenarios are worth considering because the relevant signal does not need to be a force or displacement, but could instead be a transfer of energy from the CNB into the detector. Advances in bolometer technology have led to devices capable of measuring extremely tiny energy fluxes \citep{Kenyon06, Karasik07, Bradford10, Beyer11}, so such energy transfers could potentially be easier to detect than tiny accelerations or rare nuclear changes. The magnitude of the relevant energy fluxes has only been computed for a few specific cases (e.g. when the neutrino excites phonons in a solid medium \cite{Stodolsky89, FW95}), and so it is worth exploring whether other systems could experience larger and more detectable energy changes due to their interactions with the CNB.

The report examines how efficiently energy can be transferred from cosmological neutrinos into several different potential detection systems. We begin with a review of low-energy neutrino scattering and develop a formalism for quantifying the energy flux carried into a target system by the CNB. These calculations suggest that under ideal circumstances the energy carried by cosmological neutrinos could in principle be detectable with reasonable extensions of current bolometric technologies. Next, we consider specific systems that extract energy from the CNB with various efficiencies. Consistent with previous analyses, we find that an isolated mass at rest or a mass in a static potential well cannot extract a detectable amount of energy from low-energy neutrinos. However we also find that other systems, such as a moving mass, a mass attached to a charge embedded in a magnetic field, or a mass attached to a motion-sensitive circuit, are much more efficient at extracting energy from cosmological neutrinos. Indeed, a mass coupled to a motion-sensitive circuit may even be sufficiently efficient at extracting energy from cosmological neutrinos to serve as a basis for a viable CNB detector.

\section{Low-energy neutrino scattering}

A cosmological neutrino can interact with an arbitrary laboratory system in a large number of ways, so a completely generic analysis of low-energy neutrino scattering is next to impossible. Hence this study will only consider potential neutrino detection systems where (1) the neutrino is not captured by the target system, (2) the neutrino interacts only with a dense object  (here called the ``antenna mass'' of the target system) that is much smaller than the neutrino's de Broglie wavelength, and (3) the interaction only induces motion in that object's center-of-mass, and does not excite any internal modes of the object.

The first condition is not a major constraint, since the extremely low-energy cosmological neutrinos are very inefficient at exciting nuclear reactions, and the most promising capture reactions are those already examined in the context of the beta-decay experiments mentioned above \citep{Kaboth10}. The second and third conditions, by contrast, are more restrictive.  For example, they exclude scenarios  where the neutrino directly excites phonons in a solid \citep{Stodolsky89, FW95}. 
Still, the above class of detection schemes includes not only the well-studied case of an isolated antenna mass, but also more complex (and less explored) systems where the mass is trapped in a potential well or coupled to electromagnetic fields.  Pursuant to criterion (2), the neutrino is assumed not to interact directly with the components of the target system responsible for producing these fields or forces. This can easily be achieved if the antenna mass is much denser than any other part of the system (e.g. the antenna mass is made of lead while the other components are made of aluminum). 

Previous investigations focusing on the recoil motion of an isolated mass have demonstrated that the neutrino-induced acceleration and kinetic energy  are too small to be detectable \citep{CM82, Shv82, Stodolsky89, FW95, Duda01, Ringwald09}. However, for the broader class of systems considered here  the energy transferred into the target system by the neutrino interaction does not necessarily equal the kinetic energy of the object's recoil motion. Hence we require more general expressions for how often cosmological neutrinos will interact with the target system and how much energy the neutrinos will impart into the detector. 
From standard time-dependent perturbation theory, the interaction rate between the neutrinos and the target system can be expressed as:
\begin{equation}
R=2\pi\delta(\mathcal{E}_i-\mathcal{E}_f)\left|\langle I|\mathcal{H}'|F\rangle \right|^2
\end{equation}
where $\mathcal{E}_i$ and $\mathcal{E}_f$ are the total
energies of the initial and final energy eigenstates of the combined system 
(neutrino plus target), $|I\rangle$ and $|F\rangle$ are the spatial parts of the same eigenstates, and $\mathcal{H}'$ is the interaction Hamiltonian between the neutrino and the target system. Furthermore, if the neutrinos are not captured by the target (cf. condition 1 above) and the initial and final states of the neutrinos have energies $E_i$ and $E_f$, respectively, then the rate at which energy is transferred into the target system by the neutrinos is:
\begin{equation}
P=(E_i-E_f)R=(E_i-E_f)2\pi\delta(\mathcal{E}_i-\mathcal{E}_f)\left|\langle I|\mathcal{H}'|F\rangle \right|^2
\end{equation}

These expressions for $R$ and $P$ can now be evaluated assuming that the neutrinos only interact with a mass containing $N$ fermions packed into a region that is much smaller than the neutrinos' de Broglie wavelength (cf. condition 2 above) and that the internal modes of the mass are not excited by the interaction (cf. condition 3 above). In this case, the interaction Hamiltonian can be approximated by the following simple form:
\begin{equation}
\mathcal{H}'=-\xi c_V G_F N \delta({\bf x}_\nu-{\bf x}_A)
\end{equation}
where $G_F=10^{-5}$ GeV$^{-2}$ is the Fermi constant, $c_V$ is the vector amplitude factor that depends on the fermion content in the antenna, $\xi$ is a numerical  constant that ranges between $\sqrt{2}$ for relativistic neutrinos and $1/\sqrt{2}$  for  non-relativistic Dirac neutrinos \citep{FW95}, and ${\bf x}_\nu$ and ${\bf x}_A$ are the spatial coordinates of the neutrino and the antenna mass, respectively.

Regardless of the target-system's structure,  the initial and final states of the neutrino will correspond to those of free particles. For the sake of clarity, the following calculations assume a monochromatic, unidirectional flux of neutrinos onto the target. Hence we can write the initial and final states of the combined system as $|I\rangle=\psi^{\nu}_{i}|i\rangle$ and $|F\rangle=\psi^{\nu}_{f}|f\rangle$, where $|i\rangle$ and $|f\rangle$ are the spatial components of the initial and final eigenstates of the  target system, while $\psi^{\nu}_{i}$ and $\psi^{\nu}_{f}$ are the
initial and final states of the neutrinos:
\begin{equation}
\psi^{\nu}_{i,f}=\mathcal{N}_\nu \exp(i{\bf p}_{i,f} \cdot {\bf x}_\nu)
\end{equation}
where ${\bf p}_{i,f}$ are the incoming and outgoing neutrino momenta and the normalization constant $\mathcal{N}_\nu$  is set such that $|\mathcal{N}_\nu|^2=1/\mathcal{V}$, where $\mathcal{V}$ is a  quantization volume that can be expressed in terms of either the number density of incident neutrinos $n_i$ or a differential element of the outgoing momenta ${\bf p}_f$: 
\begin{equation}
|\mathcal{N}_\nu|^2=1/\mathcal{V}=n_i=\frac{d^3{\bf p}_f}{(2\pi)^3}.
\end{equation}

Inserting these expressions into the above equations for $R$ and $P$, and integrating over ${\bf x}_\nu$ to eliminate the delta function in the Hamiltonian gives the following expressions for the interaction and energy transfer rates:
\begin{equation}
R=2\pi \xi^2c_V^2G_F^2N^2 n_i\delta(\mathcal{E}_i-\mathcal{E}_f)
\left|\langle i|e^{-i{\bf \delta p \cdot x}_A}|f\rangle\right|^2\frac{d^3{\bf p}_f}{(2\pi)^3},
\label{R3}
\end{equation}
\begin{equation}
P=2\pi\xi^2c_V^2G_F^2N^2 n_i(E_i-E_f)\delta(\mathcal{E}_i-\mathcal{E}_f)
\left|\langle i|e^{-i{\bf \delta p\cdot x}_A}|f\rangle\right|^2\frac{d^3{\bf p}_f}{(2\pi)^3}
\label{P3}
\end{equation}
where ${\bf \delta p}={\bf p}_i-{\bf p}_f$ is the change in the neutrino's momentum during the interaction.

The above rates are for an interaction that yields a neutrino with a specific momentum ${\bf p}_f$. The total interaction and energy transfer rates are obtained by integrating over all possible outgoing momenta, yielding the following expressions:
\begin{equation}
\mathcal{R}=2\pi \xi^2c_V^2G_F^2N^2n_i\int\delta(\mathcal{E}_i-\mathcal{E}_f)
\left|\langle i|e^{-i{\bf \delta p\cdot x}_A}|f\rangle\right|^2\frac{d^3{\bf p}_f}{(2\pi)^3}
\label{R4}
\end{equation}
\begin{equation}
\mathcal{P}=
2\pi\xi^2c_V^2G_F^2N^2 n_i\int(E_i-E_f)\delta(\mathcal{E}_i-\mathcal{E}_f)
\left|\langle i|e^{-i{\bf \delta p\cdot x}_A}|f\rangle\right|^2\frac{d^3{\bf p}_f}{(2\pi)^3}
\label{P4}
\end{equation}
These expressions may be simplified slightly by recognizing that  the differential element: 
\begin{equation}
d^3{\bf p}_f=p_f^2dp_fd\Omega=p_fE_fdE_fd\Omega
\end{equation}
where $p_f=|{\bf p}_f|$ and the second equality follows from the standard relation $E_f^2=p_f^2+m_\nu^2$. We may therefore integrate over $E_f$ to eliminate
the energy-conserving delta function, leaving only the angular integral:
\begin{equation}
\mathcal{R}=\xi^2c_V^2G_F^2N^2n_i\frac{1}{4\pi^2}
\int p_fE_f\left|\langle i|e^{-i{\bf \delta p\cdot x}_A}|f\rangle\right|^2d\Omega
\label{R5}
\end{equation}
\begin{equation}
\mathcal{P}=
\xi^2c_V^2G_F^2N^2 n_i\frac{1}{4\pi^2}
\int p_fE_f(E_i-E_f)\left|\langle i|e^{-i{\bf \delta p\cdot x}_A}|f\rangle\right|^2d\Omega
\label{P5}
\end{equation}
Finally, these expressions can be re-written in the following forms:
\begin{equation}
\mathcal{R}=\frac{\xi^2c_V^2}{\pi}G_F^2N^2n_ip_iE_i \mathcal{F^R}
\label{Rexp}
\end{equation}
\begin{equation}
\mathcal{P}=\frac{\xi^2c_V^2}{\pi}G_F^2N^2 n_ip_i^3\mathcal{F^P}
\label{Pexp}
\end{equation}
Where $\mathcal{F_R}$ and $\mathcal{F_P}$ are the dimensionless efficiency factors:
\begin{equation}
\mathcal{F^R}=\frac{1}{4\pi}
\int \frac{p_fE_f}{p_iE_i}\left|\langle i|e^{-i{\bf \delta p\cdot x}_A}|f\rangle\right|^2d\Omega
\label{FR}
\end{equation}
\begin{equation}
\mathcal{F^P}=\frac{1}{4\pi}
\int \frac{p_fE_f(E_i-E_f)}{p_i^3}\left|\langle i|e^{-i{\bf \delta p\cdot x}_A}|f\rangle\right|^2d\Omega.
\label{FP}
\end{equation}

The above formulation of the interaction and the energy transfer rates has the useful property that  the target-dependent factors have now been isolated into the efficiency factors. Furthermore, provided $E_f<E_i$, neither efficiency factor can exceed unity, so $\mathcal{F^P} =\mathcal{F^R}=1$  represents an ideal cosmological neutrino detector. Such a detector not only has  the highest possible neutrino interaction rate, but also extracts all of the kinetic energy avialible from each scattered neutrino. Thus we can use this ideal case to establish whether {\it any} system of this type could ever yield a detectable signal.

Real cosmological neutrinos have a range of energies and approach the target from all directions, so a precise estimate of the relevant rates  would require integrating the above expressions  for $\mathcal{R}$ and $\mathcal{P}$ over the appropriate distribution functions. While we will consider the variability in the neutrinos' approach directions as appropriate below, accounting for the finite range of neutrino energies would just complicate the expressions and this level of precision is not needed for the order-of-magnitude calculations presented in this initial study. Instead we will simply insert ``typical'' values for cosmological neutrinos into Equations~\ref{Rexp} and~\ref{Pexp}. For the sake of simplicity, we assume massless cosmological neutrinos, so $p_i=E_i \simeq 10^{-4}$ eV, and the local number density $n_i\simeq 100$/cm$^3$. Also, since the antenna mass would best be constructed of a dense metal like lead (with $\sim 10^{25}$ fermions per cubic centimeter), and it must also be smaller than the $\sim1$ cm de Broglie wavelength of these neutrinos, a reasonable value for the total number of fermions in the antenna mass is $N=10^{24}$. Inserting these numbers into Equations~\ref{Rexp} and~\ref{Pexp} gives: 
\begin{equation}
\mathcal{R}=
(10^{-4}/s)\left(\frac{\xi^2c_V^2}{\pi}\right)
\left(\frac{N}{10^{24}}\right)^2
\left(\frac{n_i}{100/cm^3}\right)
\frac{p_iE_i}{(10^{-4} eV)^2}\mathcal{F}^\mathcal{R}
\label{Rel}
\end{equation}
\begin{equation}
\mathcal{P}=
(2\times10^{-27} W)\left(\frac{\xi^2c_V^2}{\pi}\right)
\left(\frac{N}{10^{24}}\right)^2
\left(\frac{n_i}{100/cm^3}\right)
\frac{p_i^3}{(10^{-4} eV)^3}\mathcal{F^P}
\label{Pel}
\end{equation}
Thus a system with $\mathcal{F^R}\simeq 1$ would interact with cosmological neutrinos once every few hours, and if $\mathcal{F^P}\simeq 1$, the target system could extract $10^{-27}$ Watts from the Cosmic Neutrino Background. 

While $10^{-27}$ Watts is not much power, it could be within the reach of current technologies. Modern bolometric  detectors are now approaching sensitivities of order a few times $10^{-19}W/\sqrt{Hz}$ \citep{Kenyon06, Karasik07, Bradford10, Beyer11}. If such devices could be coupled to targets with $\mathcal{F^P}\simeq 1$, then the above power flux could be detected in $\sim10^8$ detector-years, or in a single  year with  $\sim10^8$ detectors. This number, while large, does not necessarily correspond to an impossibly large instrument. If each individual antenna mass is less than $\sim$1 cm$^3$ in size, the entire three-dimensional array of $10^8$ detectors could in principle fit within a region 10 meters across. Hence it may be possible to construct an array with sufficient raw sensitivity to detect cosmological neutrinos if we can find an ``efficient'' target system with $\mathcal{F^P}$ of order unity.

\section{Efficiencies of specific detector systems}

The calculations in the previous section reveal that a system with $\mathcal{F^P} \sim 1$ could be able to extract a detectable amount of energy from cosmological neutrinos. This might appear to contradict previous analyses which demonstrated that the mechanical forces produced by cosmological neutrinos are undetectable \citep{CM82, Shv82, Stodolsky89, FW95, Duda01, Ringwald09}. However, as demonstrated below, the scenarios considered in these earlier works yield a $\mathcal{F^P} \sim {E_i/M} << 1$, which means these systems can only extract an undetectably small amount of energy from cosmological neutrinos, consistent with the published calculations.  Fortunately, a careful consideration of these detectors' limitations allows us to identify detection schemes that could extract energy much more efficiently from the Cosmic Neutrino Background. 

The following sections consider a series of model detector systems. First, we examine the case of free antenna mass initially at rest, and recover the well-known result that the mechanical forces generated by cosmological neutrinos are too small to detect. Next, we consider a mass trapped in a static potential well, and show that such a system is not significantly more efficient at extracting energy from cosmological neutrinos than a free mass. We then consider systems where the mass is moving with respect to the lab frame or coupled to a charge embedded in a magnetic field, and demonstrate that such systems can be much more efficient neutrino detector than a free mass. Finally, we describe a system composed of a dense mass coupled to a motion-sensitive circuit that may be able to achieve the desired $\mathcal{F^P} \sim 1$. Note that in all these discussions the neutrino momenta $p_i$ and $p_f$ are always measured relative to the laboratory frame.

\subsection{Free mass at rest}

If the target mass is entirely free, then the initial and final states of the target system are those of free particles:
\begin{equation}
|i,f\rangle=\mathcal{N}_A \exp(i{\bf P}_{i,f}\cdot {\bf x}_A)
\end{equation}
where ${\bf P}_{i,f}$ are the initial and final momenta of the target mass in the laboratory frame, and the normalization factor $\mathcal{N}_A$ can again either be expressed in terms of a quantization volume $\mathcal{V}$ or as a differential momentum element. In this case the matrix element reduces to a momentum-conserving delta function, which is eliminated by integrating over all possible outgoing antenna momenta, leaving the following efficiency factors:
\begin{equation}
\mathcal{F}^\mathcal{R}_{free}=\frac{1}{4\pi}\int \frac{p_fE_f}{p_iE_i} d\Omega
\label{FRfree}
\end{equation}
\begin{equation}
\mathcal{F}^\mathcal{P}_{free}=\frac{1}{4\pi}\int \frac{p_fE_f(E_i-E_f)}{p_i^3} d\Omega
\end{equation}
If we further stipulate that the mass is initially at rest in the lab frame (i.e. $P_i=0$), then conservation of energy and momentum requires that $E_i - E_f ={\delta p^2}/{M}$, where $\delta p^2=|\delta {\bf p}|^2=p_i^2+p_f^2-2p_ip_f\cos\theta$ and $\theta$ is the scattering angle in the center-of-mass frame. Thus we can re-express the energy-transfer efficiency factor as:
\begin{equation}
\mathcal{F}^\mathcal{P}_{free}=\frac{1}{4\pi}\int \frac{p_fE_f}{p_i^2}\frac{\delta p}{p_i}\left(\frac{\delta p}{M}\right) d\Omega
\label{FPfree}
\end{equation}
For any reasonable antenna mass $M$ will be much larger than $E_i$, $p_i$ or $\delta p$.  In this limit, $E_i - E_f \simeq ({p_i^2}/{M})(1-cos\theta)$, so $E_i-E_f<<E_i$ and $p_i-p_f<<p_i$, and the above integrals become:
\begin{equation}
\mathcal{F}^\mathcal{R}_{free}=1,
\label{FRel}
\end{equation}
\begin{equation}
\mathcal{F}^\mathcal{P}_{free}= \frac{E_i}{M}.
\label{FPel}
\end{equation}

Since $\mathcal{F}^\mathcal{R}$ is unity, free masses can interact with cosmological neutrinos once every few hours, so the interaction rates themselves are not necessarily a major obstacle for detecting cosmological neutrinos. Instead, the primary issue is that $\mathcal{F^P} = E_i/M$ is very small. The mass of the target can be written as $M=N\mu$, where $\mu$ is the mass of the relevant fermions. Even in the best (and least realistic) case of  a pure electron target with $\mu =511$ keV, the ratio $E_i/M\sim 2*10^{-34}(N/10^{24})^{-1}$ is extremely small, and the power transfer rate is correspondingly feeble:
\begin{equation}
\mathcal{P}_{free}=
(4*10^{-59} W)\left(\frac{\xi^2c_V^2}{\pi}\right)
\left(\frac{N}{10^{24}}\right)
\left(\frac{n_i}{100/cm^3}\right)
\frac{p_i^3E_i}{(10^{-4} eV)^4}\left(\frac{511 keV}{\mu}\right)
\label{Pelw}
\end{equation}
This power is far too low to be detected with any reasonable technology, consistent with previous analyses \citep{CM82, Stodolsky89, FW95}. Thus a system which could yield a detectable signal from cosmological neutrinos would need to extract energy from the neutrinos much more efficiently than a free mass at rest. 


\subsection{Mass trapped in a static potential}

Since free masses  cannot efficiently extract energy from the cosmological neutrinos, we must consider more complex target systems where the motion of the antenna mass has a non-trivial spectrum of excited states. One simple example of such a system consists of  an antenna mass trapped in a potential well. In this scenario,  the neutrino's interaction with the mass excites transitions between eigenstates of the potential, enabling the target system to capture a significant fraction of the incoming neutrino's energy. Unfortunately, it turns out that masses trapped in fixed potentials are not significantly more efficient neutrino detectors than free masses. 

Consider a mass initially in the ground state of the potential well $|g\rangle$ with energy $\epsilon_g$, which the neutrino will excite into a state $|e\rangle$ with energy $\epsilon_e=\epsilon_g+\delta\epsilon$. Note that $E_i-E_f=\epsilon_e-\epsilon_g$. The energy transfer rate efficiency factor  is therefore:

\begin{equation}
\mathcal{F}^\mathcal{P}_{fixed}=\frac{1}{4\pi}
\int \frac{p_fE_f(\epsilon_e-\epsilon_g)}{p_i^3}\left|\langle g|e^{-i{\bf \delta p\cdot x}_A}|e\rangle\right|^2d\Omega.
\label{FPb}
\end{equation}

So long as $\delta\epsilon>0$, this expression can be re-written in the following form:
\begin{equation}
\mathcal{F}^\mathcal{P}_{fixed}=\frac{1}{4\pi}
\int \frac{p_fE_f}{p_i^3}\frac{1}{\delta\epsilon}
\left|\langle g|\epsilon_g e^{-i{\bf \delta p\cdot x}_A}-e^{-i{\bf \delta p\cdot x}_A}\epsilon_e|e\rangle\right|^2d\Omega.
\label{FPb2}
\end{equation}
The states $|g\rangle$ and $|e\rangle$ are eigenstates of the Hamiltonian  for the bound mass:
\begin{equation}
\mathcal{H}_b = -\frac{1}{2M}\hat{P}_A^2+V({\bf x}_{A})
\label{Hb}
 \end{equation}
 where $\hat{P}_A$ is the momentum operator for the antenna mass and $V({\bf x}_A)$ is the trapping potential. Thus $\mathcal{H}_b$ can replace  $\epsilon_g$ and $\epsilon_e$ in the above expression. So long as the potential is only a function of the mass position ${\bf x}_{A}$, it will commute with $e^{-i{\bf \delta p\cdot x}_A}$ and thus cancel out of the expression, leaving:
\begin{equation}
\mathcal{F}^\mathcal{P}_{fixed}=\frac{1}{4\pi}
\int \frac{p_fE_f}{p_i^3}\frac{1}{4M^2\delta\epsilon}
\left|\langle g|\hat{P}_A^2 e^{-i{\bf \delta p\cdot x}_A}-e^{-i{\bf \delta p\cdot x}_A}\hat{P}_A^2|e\rangle\right|^2d\Omega.
\label{FPb3}
\end{equation} 
Using the standard commutation rules, this expression reduces to:
\begin{equation}
\mathcal{F}^\mathcal{P}_{fixed}=\frac{1}{4\pi}
\int \frac{p_fE_f}{p_i^3}\frac{1}{4M^2\delta\epsilon}
\left|\delta p^2\langle g|e^{-i{\bf \delta p\cdot x}_A}|e\rangle-\langle g|e^{-i{\bf \delta p\cdot x}_A} (i{\bf \delta p \cdot \hat{\bf P}}_A)|e\rangle\right|^2d\Omega.
\label{FPb4}
\end{equation} 
Now define ${\bf \delta k}={\bf \delta p}/\delta p$ and ${\bf \hat{K}}_A={\bf \hat{P}}_A/\sqrt{M\delta\epsilon}$. The parameter ${\bf \delta k}$ is just the unit vector pointing along the direction of ${\bf \delta p}$, while the unitless operator ${\bf \hat{K}}_A$ 
is a linear combination of raising and lowering operators where the coefficients on these operators are of order unity. In terms of these parameters, the expression becomes:
\begin{equation}
\mathcal{F}^\mathcal{P}_{fixed}=\frac{1}{4\pi}
\int \frac{p_fE_f}{p_i^3}\frac{1}{4M^2\delta\epsilon}
\left|\delta p^2\langle g|e^{-i{\bf \delta p\cdot x}_A}|e\rangle-
\delta p\sqrt{M\delta\epsilon}\langle g|e^{-i{\bf \delta p\cdot x}_A} (i{\bf \delta k \cdot \hat{\bf K}}_A)|e\rangle\right|^2d\Omega.
\label{FPb5}
\end{equation} 
Rearranging terms and pulling out a leading factor of $p_i/M$, gives:
\begin{equation}
\mathcal{F}^\mathcal{P}_{fixed}=\frac{p_i}{M} \left[\frac{1}{4\pi}
\int \frac{p_fE_f}{p_i^2}\left(\frac{\delta p}{2p_i}\right)^2
\left|\sqrt{\frac{\delta p^2}{M\delta\epsilon}}\langle g|e^{-i{\bf \delta p\cdot x}_A}|e\rangle-
\langle g|e^{-i{\bf \delta p\cdot x}_A} (i{\bf \delta k \cdot \hat{\bf K}}_A)|e\rangle\right|^2d\Omega.\right]
\label{FPb6}
\end{equation} 
In order for the efficiency factor to be of order unity,  we need the term in brackets to be of order $M/p_i>>1$, but this is impossible. The  factors of $p_fE_f/p_i^2$ and $\delta p/2p_i$ must both be less than unity if the neutrino is to donate energy to the target system. Furthermore, the $\delta \epsilon$ for a bound particle cannot be less than its value for a free particle  $\delta p^2/2M$, so $\delta p^2/M\delta\epsilon$ also cannot exceed $\sqrt{2}$. The first matrix element cannot possibly exceed unity because of how the states are normalized. Finally, the above definition of $\hat{\bf K}_A$ should prevent the second term from being much larger than 1 (This is certainly true for simple potentials such as square wells and harmonic oscillators, but a formal proof that it also applies to more complex potentials is beyond the scope of this report). Thus  $\mathcal{F}^\mathcal{P}_{fixed}$ can never be much larger than $p_i/M$, and one cannot construct an efficient neutrino detector from isolated masses trapped in static potential wells. 

\subsection{Free mass in motion}

At first, the above calculations would appear to suggest that $\mathcal{F^P}$  must always be of order $E_i/M$. However, much more efficient systems are possible if we critically examine the assumptions behind these computations. For example, let us return to the case of a free mass, but instead of assuming the mass starts at rest, have the mass initially moving at a finite velocity relative to the laboratory frame, so the initial state of the mass has a finite initial momentum ${\bf P}_i$.  In this case, the relevant matrix element still corresponds to a momentum-conserving delta function, but if $P_i>>p_i$, then conservation of energy and momentum requires that $E_i-E_f={\bf \delta p \cdot P}_i/M$. Thus the momentum impulse $\delta p$ can potentially  produce much larger changes in the kinetic energy of a moving mass than it can for a stationary one.

For any realistic detector system $P_i/M<<1$, so $E_i-E_f<<E_i$ and the relevant efficiency factors become:
\begin{equation}
\mathcal{F}^\mathcal{R}_{move} \sim 1,
\label{FRmov}
\end{equation}
\begin{equation}
\mathcal{F}^\mathcal{P}_{move} \sim \frac{P_i}{M}(\bf{k}_i\cdot \bf{K}_i).
\label{FPmov}
\end{equation}
where ${\bf k}_i$ and ${\bf K}_i$ are unit vectors aligned with the vectors $ {\bf p}_i$ and ${\bf P}_i$, respectively. Note that $P_i/M = v_i/c$, where $v_i$ is the initial speed of the target mass

If the neutrinos were truly unidirectional, then we could make $({\bf k}_i\cdot {\bf K}_i)=1$ by ensuring the mass moves in the same direction as the incident neutrinos. Of course, real cosmological neutrinos will approach the target from all directions. If the incident neutrino flux were perfectly isotropic, then $\bf{k}_i\cdot \bf{K}_i$ would average to precisely zero. In practice, the incident neutrino flux is not exactly isotropic because the solar-system's peculiar velocity $v_p$ relative to the mean Hubble flow produces a detectable dipole variation in both the CMB and the CNB. In this situation the average $(\bf{k}_i\cdot \bf{K}_i)$ in the laboratory (solar system) frame will be of order $v_p/c$. Based on observations of the CMB, $v_p/c \simeq 0.001$ \cite{Hinshaw09}, and for the sake of argument, we may imagine that the mass is initially moving at a speed of a few centimeters per second towards the peak of the CMB dipole. In that case, the relevant efficiency factor becomes:
\begin{equation}
\mathcal{F}^\mathcal{P}_{move} \sim 3\times10^{-14} \left(\frac{v_i}{ 1 cm/s}\right)\left(\frac{v_p/c}{0.0001}\right)\label{FPmov}
\end{equation}
Since $\mathcal{F^{P}}$ is still much less than unity, moving masses are unlikely to serve as practical cosmological neutrino detectors. However, this calculation also shows that a moving mass is many orders of magnitude more efficient than the previous two systems. Thus systems with $\mathcal{F^P}>>E_i/M$ do exist, which offers hope that efficient cosmological neutrino detectors may be possible.

\subsection{Mass attached to a charged object}

Systems with initially moving masses are not the only ones that can achieve  $\mathcal{F^P}>>E_i/M$, and additional potential neutrino detection systems can be found by examining the assumptions behind the above calculations for the mass trapped in the static potential well. Specifically, that prior analysis assumed that the Hamiltonian of the target system could be expressed in terms of a fixed potential that depends only on the mass' spatial coordinates (see Equation~\ref{Hb}). However, there are also systems with Hamiltonians that are more complex functions of  momentum.  For example, the Hamiltonian of a charged particle coupled to an electromagnetic field is
\begin{equation}
\mathcal{H}_{EM}=\frac{1}{2M}\left(\hat{\bf P}_A-q{\bf A}\right)^2+q\Phi({\bf x}_A)
\end{equation}
where $q$ is the particle's charge, while $\Phi$ and ${\bf A}$ are the scalar and vector potentials of the electromagnetic field. So long as ${\bf A}\ne 0$ , this Hamiltonian will have momentum-dependent terms that are qualitatively distinct from those associated with a mass trapped in a fixed potential well. These terms do not commute with $e^{-i{\bf \delta p\cdot x}_A}$ and therefore could generate larger values for $\mathcal{F^P}$.

\begin{figure}
\resizebox{6in}{!}{\includegraphics{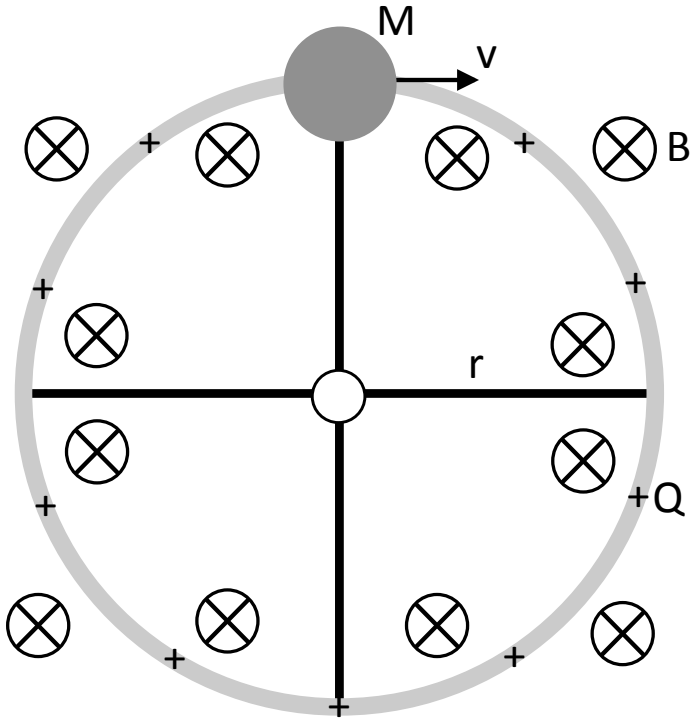}}
\caption{Diagram of a potential neutrino-detection system consisting of a mass $M$ attached to a charged ring embedded in a magnetic field. Note that the mass' motion will cause the ring to rotate about its own axis.}
\label{loop}
\end{figure}

For example, consider the system illustrated in Figure~\ref{loop}, where the antenna mass is attached to a rigid ring of radius $r$ with negligible mass and carrying an electric charge $Q$. This ring is embedded in a constant magnetic field $B$, and is held such that it can only rotate along an axis aligned with that field. In this scenario, the constraints on the system's motion allow the Hamiltonian to be written in the following form:
\begin{equation}
\mathcal{H}_{charge}=\frac{1}{2M}\left(\hat{L}_\phi/r-QBr/c\right)^2
\end{equation}
where $\hat{L}_\phi$ is the angular momentum of the mass (and ring) moving in a circular path around the ring's axis.  Furthermore, the initial and final states of the target mass are given by the following expression: 
\begin{equation}
|i,f\rangle=\frac{1}{\sqrt{2\pi}}\exp(iL_{i,f}\phi)
\end{equation}
where $\phi$ is the angular coordinate of the mass. For the sake of simplicity, let us assume that the radius $r$ is much larger than the neutrino's de Broglie wavelength and the scale of the neutrino's wavepacket. In this limit, we may approximate the above states as the initial and final states of a particle that is constrained to move in one spatial direction (i.e. the spacing between rotational energy levels is much less than the energy imparted by the neutrino collision). The matrix element $|\langle i|e^{-1\delta{\bf p}\cdot{\bf x}_A}|f\rangle|^2$ can then be reduced to a momentum-conserving delta-function that is eliminated by integrating over all outgoing antenna mass momenta. If we also assume that initially the mass is nearly at rest, then conservation of energy and angular momentum requires that $E_i-E_f = \delta p(\delta p/M -  {\rm sign}(L_f-L_i)QBr/Mc)$. Note that since $\delta p>0$, the sign of the second term in this equation depends on the direction in which the mass moves in response to the neutrino collision.
The relevant efficiency factors can then be written in the following forms:
\begin{equation}
\mathcal{F}^\mathcal{R}_{charge}=\frac{1}{4\pi}\int \frac{p_fE_f}{p_iE_i} d\Omega
\end{equation}
\begin{equation}
\mathcal{F}^\mathcal{P}_{charge}=\frac{1}{4\pi}\int \frac{p_fE_f}{p_i^2}\frac{\delta p}{p_i}\left(\frac{\delta p}{M}-{\rm sign}(L_f-L_i)\frac{QBr}{Mc}\right) d\Omega
\end{equation}
The above expression for the rate efficiency factor is basically the same as that for the free mass (see Equation~\ref{FRfree} above), so  the interaction between the charge and magnetic field does not directly influence the rate of neutrino interactions. By contrast, the energy transfer factor has a new term proportional to $QBr/Mc$ that did not appear in Equation~\ref{FPfree} above. This term arises because transferring momentum to the target mass does not just  increase the mass' kinetic energy, it also changes the electromagnetic energy associated with the charges' motion through the magnetic field. 

It is not hard to construct a system where the second term in $\mathcal{F}^\mathcal{P}_{charge}$ is much larger than the first one. First, realize that the charge $Q$ can be expressed as the product $C_Q\Phi_o$, where $C_Q$ is the self-capacitance of the charged ring and $\Phi_o$ is its electrostatic potential. For a ring of radius $r$, the self-capacitance will be of order $4\pi\epsilon_or$, where $\epsilon_o$ is the permittivity of free space. Hence $C_Q \simeq 10^{-11} F (r/0.1m)$. If we further assume that $\Phi_o \sim 10$V and $B \sim 1$ T, and say $M \sim 1$g (consistent with $10^{24}$ nucleons) then the relevant ratio becomes:
\begin{equation}
\frac{QBr}{Mc} = 3\times10^{-17}\left(\frac{C_Q}{10^{-11}F}\right)
\left(\frac{\Phi_o}{10 V}\right)\left(\frac{r}{0.1 m}\right)\left(\frac{B}{1 T}\right)
\left(\frac{1 g}{M}\right)
\end{equation}
This ratio is orders of magnitude larger than $\delta p/M \sim E_i/M$,  but is still much less than unity, so $E_i-E_f <<E_i$. Furthermore, because the energy change depends on whether the mass' angular momentum change is positive or negative, the energy transfer efficiency depends on the approach direction of the incident neutrinos. Thus, just like the moving mass described above, the average energy transfer efficiency factor for the nearly-isotropic CNB is reduced by a factor of $v_p/c$. Hence the above expressions for the efficiency factors are:
\begin{equation}
\mathcal{F}^\mathcal{R}_{charge}\sim 1
\end{equation}
\begin{equation}
\mathcal{F}^\mathcal{P}_{charge}\sim\frac{QBr}{Mc}\frac{v_p}{c} \sim 3\times10^{-20}\left(\frac{C_Q}{10^{-11}F}\right)
\left(\frac{\Phi_o}{10 V}\right)\left(\frac{r}{0.1 m}\right)\left(\frac{B}{1 T}\right)
\left(\frac{0.1 g}{M}\right)\left(\frac{v_p/c}{0.001}\right)
\label{FPcharge}
\end{equation}
The latter equation shows that $\mathcal{F^P}$ is still much less than unity (and the moving mass described above), so this system will not yield a truly efficient neutrino detector for  any plausible value of $\Phi_o$ or $B$. However, this calculation does demonstrate that  systems do not need to have a moving mass to achieve $\mathcal{F^P}>>E_i/M$. Furthermore, these calculations confirm that a coherently-scattered neutrino does not have to deposit the vanishingly small fraction of its initial energy into a target. 

While the above efficiency factors were computed using a quantum-mechanical formalism, most aspects of this calculation can be reproduced using classical electrodynamics. A ring of radius $r$ carrying a charge $Q$ moving at a speed $v=\delta p/M$ can be approximated as a current loop which has a magnetic dipole moment $\mu\propto Qr(\delta p/M)$. If this loop is embedded in a magnetic field, it has an energy carried by that magnetic moment is $E_{loop}=\mu B \sim QBr(\delta p/M)$. This is the same energy change that forms the dominant term in the above above expression for $E_i-E_f$. This concordance arises because the extra terms in the Hamiltonian do not explicitly depend on the position of the target mass. These terms therefore do not influence the relevant matrix elements, but only affect the energy of the target system's initial and final eigenstates. Hence we should be able to estimate the efficiencies of certain more complex electromagnetically-coupled systems from how their energy should change when the relevant part of the system moves.

\subsection{Mass coupled to a motion-sensitive circuit}

\begin{figure}
\resizebox{6in}{!}{\includegraphics{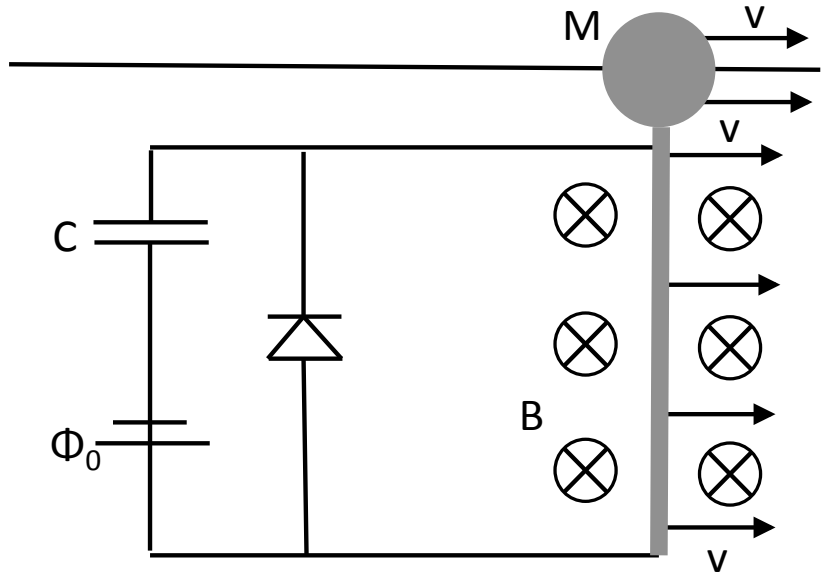}}
\caption{A possible neutrino-detector consisting of a dense antenna mass attached to a motion-sensitive circuit. Note that the motion of the mass will drag the wire (in grey) across the magnetic field.}
\label{nucirc}
\end{figure}

A simple charged system like the one described in the previous section is unlikely to yield an efficient neutrino detector because of practical limitations on the magnitudes of the object's charge and the applied magnetic field. In particular, the small self-capacitance of the ring limits the amount of charge that can be applied to it, and thus limits the energy changes associated with the system's motion. Fortunately, other electromechanical systems can have much larger energy changes induced by a moving antenna mass. 

For example, consider the system illustrated in Figure~\ref{nucirc}, where the antenna mass is constrained to move along a single linear direction, and is attached to a rigid wire of length $\ell$ that is held perpendicular to the mass' direction of motion. The wire in turn is embedded in a uniform magnetic field $B$ that is oriented perpendicular both to the mass' direction of motion and the wire's axis. Thus, if the mass moves, the wire will be drawn across the magnetic field lines, producing an electromotive force across the wire. This wire is also part of a circuit containing a battery that produces a DC voltage $\Phi_o$ and a capacitor with capacitance $C$. So long as the mass and wire segment are not moving relative to the magnetic field, the potential across the capacitor will be $\Phi_o$, and the energy stored by the capacitor will be $E_{c,0}=(1/2)C\Phi_o^2$. However, if the mass and wire move at a speed $v$, then the electromotive force due to the wire's motion through the magnetic field will produce a potential difference of $\delta \Phi = vB\ell$ across the two ends of the moving wire. In general, this potential drop could be either positive or negative depending on the wire's direction of motion. As in the previous two cases, this would reduce the efficiency of the detector to the nearly isotropic CNB by a factor of $v_p/c$. However, in this case one can insure the  $\delta \Phi$ transmitted to the capacitor has the same sign as $\Phi_o$ with an appropriate rectifier circuit (in the figure we illustrate the rectifier as a single diode that ensures $\delta \Phi >0$). Hence the potential difference across the capacitor changes from $\Phi_o$ to  $\Phi_o+\delta \Phi$ and the energy stored in the capacitor increases by $\delta E_C=C\Phi_o\delta \Phi=C\Phi_oB\ell v$. 

As with the simple charged system described in the previous section, any state  of the above system where the antenna mass has a well-defined momentum will also have a well-defined energy. Hence the relevant matrix elements are again the same as those for a mass free to move in one direction. Furthermore, assuming the target mass is initially at rest, then $E_i-E_f=\delta E_C=C\Phi_oB\ell \delta p/M$. The efficiency factor of this system is therefore given by the following expression:
\begin{equation}
\mathcal{F}^\mathcal{P}_{circuit}=\frac{1}{2\pi}\int \frac{p_fE_f}{p_i^2}\frac{\delta p}{p_i}\left(\frac{C\Phi_oB\ell c}{M}\right) d\Omega
\end{equation}
If we assume that $B$ is of order 1 T, $\Phi_o$ is of order 10 Volts, the wire is 1 meter long and the $C$ is of order 1 Farad, then we find that the above efficiency factor will be of order:
\begin{equation}
\mathcal{F}^\mathcal{P}_{circuit}\simeq 3\times10^{-5}\left(\frac{C}{1F}\right)\left(\frac{\Phi}{10 V}\right)\left(\frac{B}{1 T}\right)\left(\frac{\ell}{1m}\right)\left(\frac{1 g}{M}\right)\end{equation}
The efficiency factor of this system is therefore not much less than one. Furthermore, the efficiency of this system can easily be increased by (1) increasing the capacitance of the circuit using multiple capacitors  wired in parallel, (2) increasing the battery's voltage, and (3) replacing the straight wire with a coil that has multiple wire segments  passing through the magnetic field. These improvements could each increase $\mathcal{F^P}$ by over  an order of magnitude, hence a system similar to the one described above could act as an efficient neutrino detector, with $\mathcal{F^P}\sim1$.

\section{Conclusions}

The above analysis indicates that a compact, dense mass coupled to a suitable motion-sensitive circuit should be able to both interact with neutrinos every few hours, and extract a significant fraction of the neutrino's kinetic energy from each interaction. One of these detectors could therefore absorb of order $10^{-27}$ Watts of power from the Cosmic Neutrino Background. An array of $10^8$ such detectors could therefore absorb enough energy to be detectable with current state-of-the art bolometers. Of course, it is not yet certain that a large enough array of detectors could be constructed or made sufficiently sensitive to detect cosmological neutrinos. Furthermore, a practical neutrino detector must not just be able to detect the energy from the CNB, but also isolate that signal from other sources of excitation. Such challenges are beyond the scope of this report, but the calculations presented above indicate that a cosmological neutrino detector may not be entirely  beyond the reach of near-future technologies.


\acknowledgments I wish to thank I. Wasserman and D. Chernoff for many useful conversations. I would also like to thank the anonymous reviewers whose comments substantially improved this manuscript.


\providecommand{\href}[2]{#2}\begingroup\raggedright\endgroup

\end{document}